\documentclass{PoS}
\usepackage{graphicx}
\usepackage[latin1]{inputenc}
\usepackage{epsfig}
\psfull
\usepackage{color}
\usepackage{colortbl}
\let\normalcolor\relax
\usepackage{amsmath}
\usepackage{bbm}

\newcommand{\lrd}{\raisebox{0.09em}{$
\stackrel{\scriptstyle\leftharpoonup\hspace{-.7em}\rightharpoonup}{D}$}}
\newcommand{\ld}{\raisebox{0.09em}{$
\stackrel{\scriptstyle\leftharpoonup}{D}$}}
\newcommand{\rd}{\raisebox{0.09em}{$
\stackrel{\scriptstyle\rightharpoonup}{D}$}}

\newcommand{\ket}[1]{\left| #1 \right>}
\newcommand{\bra}[1]{\left< #1 \right|}

\newcommand{\e}{\mathrm{e}}

\newcommand{\di}{\mathrm{d}}

\newcommand{\I}{\mathrm{i}}

\newcommand{\eVdist}{\kern-0.06667em}

\newcommand{\Gev}{{\text{Ge}\eVdist\text{V\/}}}

\newcommand{\mev}{{\,\text{Me}\eVdist\text{V\/}}}
\newcommand{\gev}{{\,\text{Ge}\eVdist\text{V\/}}}

\newcommand{\tsi}{t_{\text{sink}}}
\newcommand{\tsc}{t_{\text{source}}}
\newcommand{\tsh}{t_{\text{shift}}}

\PoS{PoS(LAT2005)360}
\title{\begin{picture}(0,0)(0,0)%
\put(0,75){\makebox(0,0)[l]{\textnormal{\normalsize DESY 05-169}}}%
\put(0,60){\makebox(0,0)[l]{\textnormal{\normalsize Edinburgh 2005/10}}}%
\end{picture}%
Structure of the Pion from Full Lattice QCD
}

\ShortTitle{Structure of the Pion from Full Lattice QCD }

\author{\speaker{D.~Br\"ommel}$^{\,a,b}$, M.~Diehl$^a$, M.~G\"ockeler$^b$,
  Ph.~H\"agler$^c$, R.~Horsley$^d$, D.~Pleiter$^e$, P.E.L.~Rakow$^f$,
  A.~Sch\"afer$^b$, G.~Schierholz$^{a,e}$ and J.M.~Zanotti$^e$\\
  \llap{$^a$}Theory Group, Deutsches Elektronen-Synchrotron DESY, 22603 Hamburg,
  Germany\\
  \llap{$^b$}Institut f\"ur Theoretische Physik, Universtit\"at Regensburg,
  93040 Regensburg, Germany\\
  \llap{$^c$}Institut f\"ur Theoretische Physik T39, Physik-Department der TU
  M\"unchen,\\
   85747 Garching, Germany\\
  \llap{$^d$}School of Physics, University of Edinburgh, Edinburgh EH9 3JH, UK\\
  \llap{$^e$}John von Neumann-Institut f\"ur Computing NIC~/~DESY, 15738
  Zeuthen, Germany\\
  \llap{$^f$}Theoretical Physics Division, Dep.\ of Math.\ Sciences, University of
  Liverpool,\\
  Liverpool L69 3BX, UK\\
E-mail: \email{dirk.broemmel@desy.de}}
\author{QCDSF/UKQCD Collaboration}

\abstract{Moments of generalised parton distributions can be related to
  off-forward matrix elements of local operators. We calculate a few of the
  leading twist matrix elements for the pion on the lattice. The simulations are
  performed using two flavours of dynamical fermions and a range of pion masses
  from 550 to 1090\mev. Our lattice spacings and spatial volumes lie in the
  range \mbox{0.07--0.12~fm} and \mbox{$($1.6--2.2~fm$)^3$}, respectively. Key
  features of our investigation are the use of $O(a)$ improved Wilson fermions
  and non-perturbative renormalisation.

  We present first results for the two lowest moments of the generalised parton
  distributions of the pion and compare the pion electromagnetic form factor
  $F_{\pi}$ to experimental data. Good agreement is found between lattice data
  and experiment.}

\FullConference{XXIIIrd International Symposium on Lattice Field Theory\\
  25-30 July 2005\\
  Trinity College, Dublin, Ireland}

\begin{document}

\section{Introduction}

The pion as a near Goldstone boson is essential for chiral symmetry breaking.
It also plays an important role for theoretical models since it is the simplest
spinless meson. However, the understanding of the structure of the pion is still
limited.
On the experimental side results from electroproduction measurements, e.g.\
$e\,p \to e\,\pi^{+}\,n$, are available and provide information about the pion
electromagnetic form factor $F_{\pi}$. Parton distribution functions on the
other hand are obtained from Drell-Yan dilepton production processes,
$\pi^{\pm}\,N \to \mu^{+}\mu^{-}\,X$, and prompt photon production, i.e.\
$\pi^{\pm}\,N \to \gamma\,X$.
For some time now it has been possible to explore the structure also from first
principles using lattice QCD. Initial studies by Martinelli {\it{et al.\/}} and
Draper {\it{et al.\/}} were performed for the pion form factor and parton
distributions \cite{Martinelli:1987bh}. The main
interest recently was on the pion form factor
\cite{Bonnet:2004fr,vanderHeide:2003kh,cappos}. In this contribution we
investigate generalised parton distributions (GPDs) of the pion
\cite{Belitsky:2005qn,Diehl:2003ny,Goeke:2001tz}.

GPDs can be seen as a generalisation of parton distributions and form
factors. They contain both as limiting cases but go beyond that and include new 
information about the structure of the pion that is not yet accessible by
experiments. This work is an extension of earlier studies \cite{Best:1997qp} and
complements current efforts on the nucleon structure
\cite{Gockeler:2005cj,:2003is}.

\section{Generalised Form Factors}

GPDs parametrise off-forward matrix elements probing single quarks inside
hadrons. Hence the respective structure of hadrons can be described by GPDs. In
case of the pion, the vector operator matrix elements on the light-cone with the
corresponding GPDs $H^{q}_{\pi}$ read
\begin{equation}
  \label{eq:gpd:def}
  H^{q}_{\pi}(x,\xi,t) = \frac{1}{2\, n\cdot P}\int\frac{\di{\lambda}}{2\pi}\;
  \e^{\I \lambda n\cdot Px} \bra{\pi(p')}
  \overline{\psi}_{q}(-\frac{\lambda}{2}n)\; n\cdot\gamma\;\mathcal{U}\,
  \psi_{q}(\frac{\lambda}{2}n) \ket{\pi(p)}\,,
\end{equation}
where the kinematical variables are the average momentum of the incoming and
outgoing pion, $P^{\mu} = \frac{1}{2} (p'^{\mu}+p^{\mu})$, the momentum transfer
$\Delta^{\mu} = (p'^{\mu}-p^{\mu})$ and its invariant form $t=\Delta^2$. The
fractional longitudinal momentum transfer is $\xi = -(n\cdot\Delta)/(2n\cdot P)$ with a
light-like vector $n^{\mu}$. A Wilson line $\mathcal{U}$ is included to ensure
gauge invariance. Important properties of $H^{q}_{\pi}(x,\xi,t)$ are:
\begin{itemize}
\item%
  In the forward limit, $\Delta\to 0$, one recovers the usual parton
  distributions. We have $H^{q}_{\pi}(x,0,0)=q(x)$ for $x>0$ and
  $H^{q}_{\pi}(x,0,0)=-\overline{q}(-x)$ for $x<0$.
\vspace*{-.8ex}\item%
  The first moment in $x$ is related to the pion form factor. Starting from a
  flavour dependent GPD, the full form factor including all quark flavours can
  be be obtained using isospin relations \cite{Diehl:2003ny}. Assuming we probe
  $u$-quarks within a $\pi^{+}$ we find $\int \di x\, H^{u}_{\pi^{+}}(x,\xi,t) =
  F_{\pi}(t)$.
\vspace*{-.8ex}\item%
  Taking the Fourier transform in $\Delta_{\perp}$ results in a probabilistic
  interpretation in coordinate space. One finds $(2\pi)^{-2}\int
  \di^{2}\Delta_{\perp}\, \e^{-\I b_{\perp}\cdot\Delta_{\perp}}
  H^{q}_{\pi}(x,\xi=0,t=-\Delta^{2}_{\perp}) = q_{\pi}(x,b_{\perp})$ which is
  the probability of finding a quark $q$ with momentum fraction $x$ and impact
  parameter $b_{\perp}$ in the pion \cite{Burkardt:2002hr}.
\end{itemize}
Since Eq.~\eqref{eq:gpd:def} is a light-cone matrix element, it cannot be
calculated on the lattice directly. Instead, the operator product expansion is
applied to obtain matrix elements of local operators. These matrix elements can
then be parametrised by generalised form factors (GFFs), which are proportional
to moments of the GPDs \cite{Belitsky:2005qn,Diehl:2003ny}. The GFFs thus
provide an equivalent description of the hadron structure. The operators
$\mathcal{O}^{\{\mu \mu_1 \mu_2 \dots \mu_n\}}$ and the decomposition of their
matrix elements into GFFs corresponding to Eq. \eqref{eq:gpd:def} read (assuming
from now on that we probe $u$-quarks within a $\pi^{+}$)
\begin{multline}
 \label{eq:gff:def}
 \bra{\pi^{+}(p')} \mathcal{O}^{\{\mu \mu_1 \mu_2 \dots \mu_n\}} \ket{\pi^{+}(p)} =
  \bra{\pi^{+}(p')} \overline{u}(0)\, \gamma^{\{\mu}\, \I\lrd\/^{\mu_1}
  \I\lrd\/^{\mu_2} \dots \I\lrd\/^{\mu_n\}} \,u(0)
  \ket{\pi^{+}(p)} = \\
    2\; P^{\{\mu}P^{\mu_1} \dots P^{\mu_n\}} A_{n+1,0}(\Delta^2)
    +\; 2\sum^n_{\substack{i=1\\\textrm{odd}}} \Delta^{\{\mu}
    \Delta^{\mu_1} \dots \Delta^{\mu_i}
    P^{\mu_{i+1}} \dots P^{\mu_n\}} \,A_{n+1,i+1}(\Delta^2)\,,
\end{multline}
where $n$ labels the number of covariant derivatives $\lrd = \frac{1}{2} (\rd -
\ld)$ and $\{\dots\}$ indicates symmetrisation of indices and subtraction of
traces. The GFFs are denoted by $A_{n,i}$. This decomposition is constrained by
Lorentz invariance, parity, and time reversal, the latter requiring an even
number of momenta $\Delta$. The simplest case $n=0$ of Eq.~\eqref{eq:gff:def}
yields the pion electromagnetic form factor $\bra{\pi(p')} \mathcal{O}^{\mu}
\ket{\pi(p)} = 2\, P^{\mu} F_{\pi}(\Delta^2)$, thus we have $A_{1,0} =
F_{\pi}$.

\section{Lattice Techniques}

The calculation of the matrix elements and the extraction of GFFs in
\eqref{eq:gff:def} is done in analogy to the nucleon case. We compute pion
two- and three-point functions in order to isolate the observables in question
\cite{Gockeler:2005cj,:2003is}. Inserting
complete sets of energy eigenstates and employing translational
invariance, the three-point function takes the following form for $t<\tsi$
\begin{equation}
  \label{eq:3pt}
  C_{\pi\mathcal{O}\pi}(t,p',p) = \bra{\pi(p')} \mathcal{O}(t) \ket{\pi(p)}
  \frac{\bra{0} \eta_{\pi}(\vec{p}') \ket{\pi(p')}\bra{\pi(p)} \eta_{\pi}^{\dagger}(\vec{p})
    \ket{0}}
  {2E_{p'}\,2E_{p}}
  \text{e}^{-E_{p'}(\tsi-t)-E_{p}\,t} +\dots\,,
\end{equation}
where we omit excited states and set $\tsc=0$. Here $\eta_{\pi}(\vec{p})$ is the
interpolating field for a pion with momentum $\vec{p}$ and energy $E_{p}$ the
corresponding energy. We use both pseudo-scalar and axial vector
interpolators. The two-point function, again omitting higher energy states, has
the usual form
\begin{equation}
  \label{eq:2pt}
  C_{\pi\pi}(t,p) = \frac{\bra{0} \eta_{\pi}(\vec{p}) \ket{p} \bra{p}
    \eta_{\pi}^{\dagger}(\vec{p}) \ket{0}}
  {2E_{p}} \text{e}^{-E_{p}T/2} \cosh [ E_{p}(T/2-t) ] +\dots\,,
\end{equation}
with the time extent $T$ of the lattice. We then construct an appropriate
ratio of two- and three-point functions to eliminate the exponential time
behaviour and the overlap factors such as $\bra{0} \eta_{\pi}(\vec{p}')
\ket{\pi(p')}$ that appear in Eq.~\eqref{eq:3pt}. In doing so we avoid having to
fit the energies $E_{p}$ and the normalisation separately. We choose $\tsi =
T/2$ so that the three-point correlator is symmetric under $t \to T-t$. The
ratio that can then be used is
\begin{equation}
  \label{eq:ratio}
  \frac{\bra{p'}\mathcal{O}(t)\ket{p}}{4\sqrt{E_{p'}E_{p}}} = 
  \frac{C_{\pi\mathcal{O}\pi}(t,p',p)}{C_{\pi\pi}(\tsi,p')} \;
  \left( %
    \frac{C_{\pi\pi}(\tsi-t,p)\, C_{\pi\pi}(t,p')\,
      C_{\pi\pi}(\tsi,p')}%
    {C_{\pi\pi}(\tsi-t,p')\, C_{\pi\pi}(t,p)\,
      C_{\pi\pi}(\tsi,p)}%
  \right)^{\frac{1}{2}}\,.
\end{equation}
The l.h.s.\ now contains the matrix element we are interested in. Hence we can
extract the GFFs from matching the lattice ratio \eqref{eq:ratio} to its
continuum parametrisation Eq.~\eqref{eq:gff:def}. Using all momentum
combinations and operators available, this results in an over-constrained
fit of the $A_{n,i}(\Delta^2)$.

Contributions from excited states with energy $E'$ to the ratio \eqref{eq:ratio}
are suppressed as long as $\tsi-t \gg 1/E'$ and $t \gg 1/E'$. However, because
of the exponential decay of the pion two-point function, the signal at $t=\tsi$
for non-vanishing momenta is poor and can take negative values. This causes
problems when taking the square root. We can partly circumvent this by shifting
the two-point function, i.e.\ changing $C_{\pi\pi}(\tsi,p) \to
C_{\pi\pi}(\tsi-\tsh,p)$. This shift is compensated by an extra factor of
$(\cosh(E_{p}\, \tsh))^{-1}$ to the two-point function.

The operators we use are all constructed to be traceless and symmetric. Along
with more details like transformation and mixing properties, they are given in
\cite{Gockeler:1996mu}.

\section{Results}

Our simulation is performed with two flavours of non-perturbatively
clover-improved dynamical Wilson fermions and Wilson glue. We use operators with
up to three derivatives \cite{Gockeler:1996mu} and exploit the full
Clifford-Algebra, i.e. we calculate (pseudo-) scalar, (pseudo-) vector, and
tensor currents. The combinations involving $\gamma_{5}$ should vanish due to
the symmetry properties of \eqref{eq:gff:def} under parity and we use this as a
check for our calculations. The large number of momentum combinations for the
over-constrained fit to the GFFs is achieved by using three sink
momenta $\vec{p}'$ and 17 momentum transfers $\vec{\Delta}=\vec{p}'-\vec{p}$,
which in lattice units $\times L/2\pi$ are given by:
\begin{itemize}
\item[$\vec{p}'$:]%
  $(0,0,0)$, $(0,-1,0)$, $(-1,0,0)$,
\vspace*{-.8ex}\item[$\vec{\Delta}$\phantom{$'$}:]%
  $(0,0,0)$, $(1,0,0)$, $(1,1,0)$, $(1,1,1)$, $(2,0,0)$, $(2,1,1)$, and
  permutations w.r.t.\ the components.
\end{itemize}
A list of further parameters of our simulation can be found in Table
\ref{tab:lats}. The configurations have been generated within the QCDSF and
UKQCD collaborations.
\begin{table}[b]
  \begin{center}
  {\small
  \begin{tabular}{@{}c@{$\;$}|@{$\;$}c@{$\;$}|@{$\;$}c@{$\;$}|@{$\;$}c@{$\;$}|@{$\;$}>{\columncolor[rgb]{.95,.95,.95}[.33\tabcolsep]}c@{$\;$}|@{$\;$}>{\columncolor[rgb]{.95,.95,.95}[.33\tabcolsep]}c@{$\;$}|@{$\;$}>{\columncolor[rgb]{.95,.95,.95}[.33\tabcolsep]}c@{$\;$}|@{$\;$}>{\columncolor[rgb]{.95,.95,.95}[.33\tabcolsep]}c@{$\;$}|@{$\;$}c@{$\;$}|@{$\;$}>{\columncolor[rgb]{.95,.95,.95}[.33\tabcolsep]}c@{$\;$}|@{$\;$}>{\columncolor[rgb]{.95,.95,.95}[.33\tabcolsep]}c@{$\;$}|@{$\;$}c@{$\;$}|@{$\;$}c@{}}
    \hline\hline
    $\beta$ & \rule{0pt}{2.6ex}5.20 & 5.20 & 5.20 & \textbf{5.25} &
    \textbf{5.25} & \textbf{5.25} & \textbf{5.29} & 5.29 & \textbf{5.29} &
    \textbf{5.40} & 5.40 & 5.40 \\[.4ex]
    $\kappa$ & .13420 & .13500 & .13550 & \textbf{.13460} & \textbf{.13520} &
    \textbf{.13575} & \textbf{.13400} & .13500 & \textbf{.13550} &
    \textbf{.13500} & .13560 & .13610 \\[.4ex]
    $m_{\pi}$ [\Gev] & .94 & .777 & .578 & \textbf{.92} & \textbf{.774} &
    \textbf{.553} & \textbf{1.09} & .867 & \textbf{.716} & \textbf{.969} & .788 &
    .588\\[.4ex]
    $L$ [fm] & 1.96 & 1.68 & 1.59 & \textbf{1.69} & \textbf{1.56} &
    \textbf{2.19} & \textbf{1.66} & 1.53 & \textbf{2.16} & \textbf{1.97} & 1.88
    & 1.79\\[.4ex]
    $m_{\pi}\cdot L$ & 9.36 & 6.64 & 4.66 & \textbf{7.89} & \textbf{6.11} &
    \textbf{6.14} & \textbf{9.23} & 6.74 & \textbf{7.85} & \textbf{9.68} & 7.48
    & 5.3\\[.06ex]
    \hline\hline
  \end{tabular}
  \caption{Our set of lattices available. Highlighted columns mark the current data points.}
  \label{tab:lats}
  }\end{center}
\end{table}
The connection to physical values is done using the Sommer scale with
$r_{0}=0.5$~fm and non-perturbative renormalisation
\cite{Martinelli:1994ty}. The results have finally been converted to the
$\overline{\text{MS}}$-scheme at $\mu=2\gev$.

\begin{figure}[t!]
\begin{minipage}[c]{.48\textwidth}
    \includegraphics[height=1.25\textwidth]{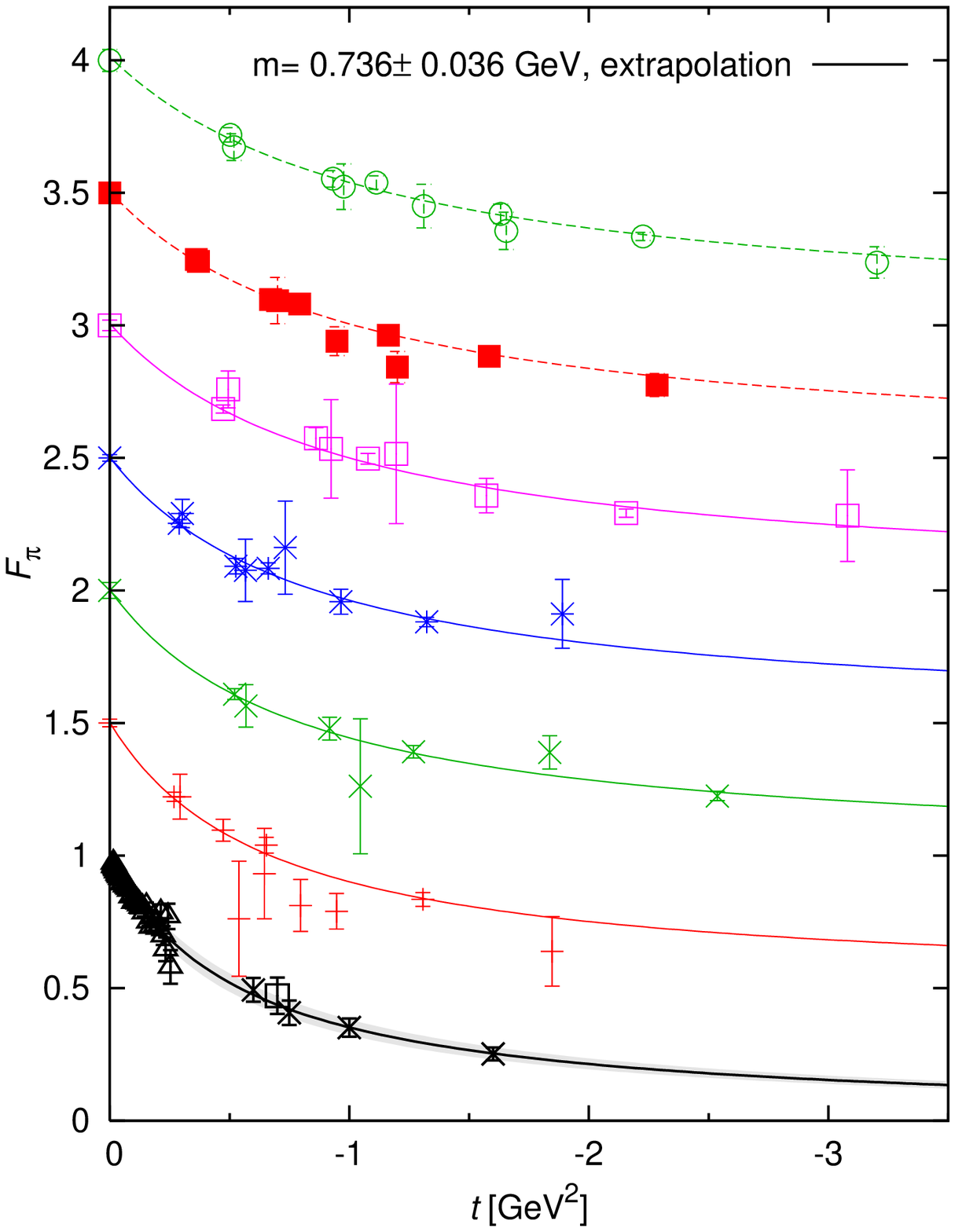}
    \caption{\label{fig:f_pi}Pion form factor $F_{\pi}(t)$ for our different
      lattices (with different offsets). The pion masses decrease from top to
      bottom with fitted monopole masses: %
      $1066(43)\mev$, $1005(18)\mev$, $993(28)\mev$, $926(24)\mev$,
      $892(32)\mev$, and $817(26)\mev$. Also included is the extrapolation to
      the physical pion mass with a comparison to experimental data (black curve
      and symbols).}
\end{minipage}
\hfill
\begin{minipage}[c]{.48\textwidth}
    \includegraphics[angle=-90,width=1.03\textwidth]{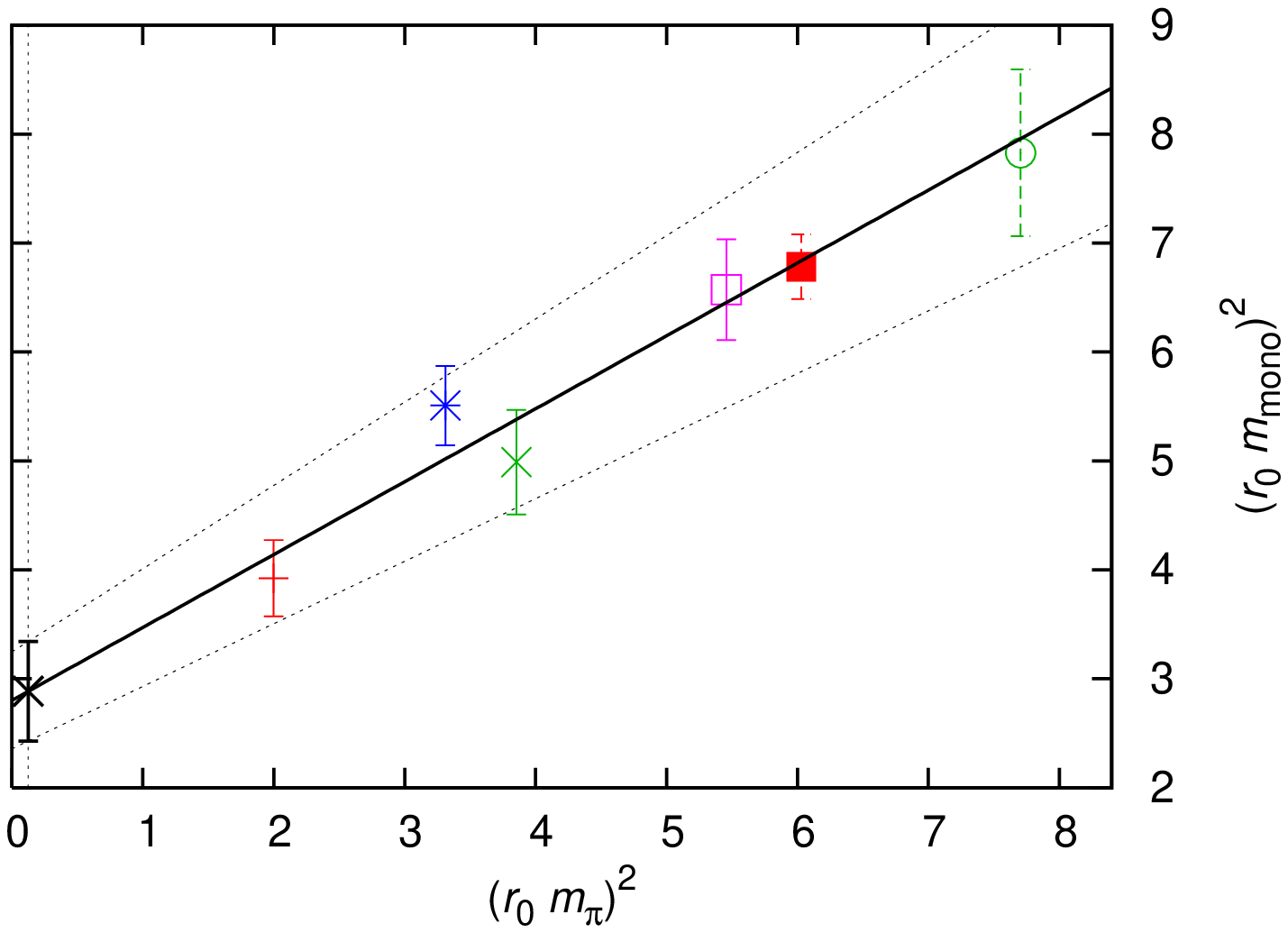}
    \caption{\label{fig:extrapol}Linear chiral extrapolation of the monopole
      masses to the physical $m_{\pi}$.}
\vspace*{-.2ex}
    \includegraphics[angle=-90,width=1.03\textwidth]{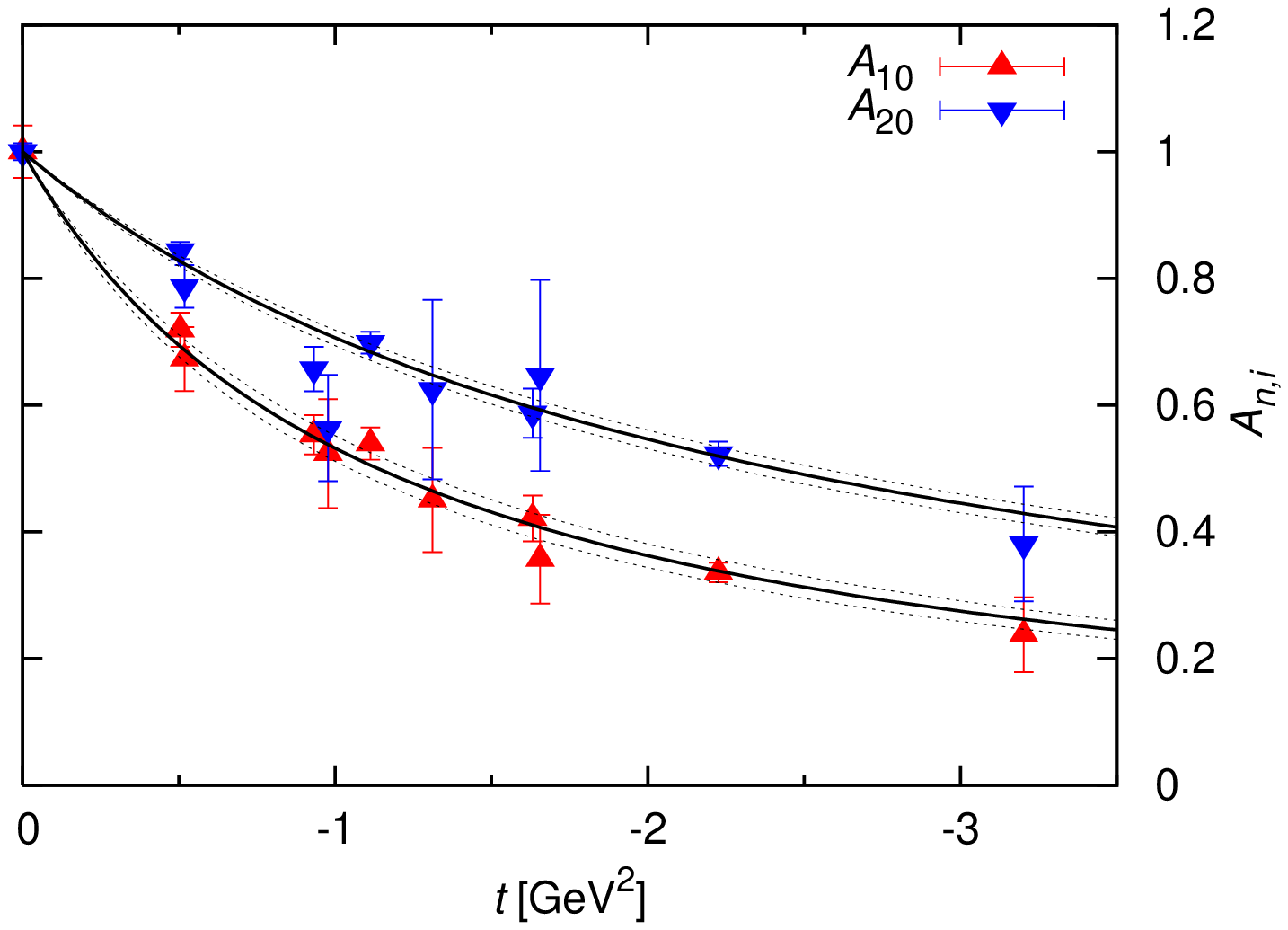}
    \caption{\label{fig:moments}First and second moment, $A_{1,0}$ and
      $A_{2,0}$, of the pion GPD [$\beta=5.29$, $\kappa=0.13400$].}
\end{minipage}
\end{figure}

Comparing both interpolating fields for the pion, we find a cleaner signal for
the axial vector current. Using this interpolator we extract values for
$A_{1,0}$ and $A_{2,0}$. Results for $A_{1,0}$, the pion electromagnetic form
factor, can be found in Fig.~\ref{fig:f_pi}. We use a monopole ansatz
$(1-t/m_{\textrm{mono}}^{2})^{-1}$ to fit our data and find that the monopole
mass decreases with decreasing pion masses as expected. A linear chiral
extrapolation of the monopole masses, shown in Fig.~\ref{fig:extrapol}, provides
a mass of $736(36)\mev$ in the chiral limit. This is in nice agreement with
experimental data, as shown by the bottom curve in Fig.~\ref{fig:f_pi}. Scaling
violations are expected to be small and will be investigated in more detail at a
later stage when the analysis is completed for all lattices.

In Fig.~\ref{fig:moments} we show $A_{1,0}$ and $A_{2,0}$ for one of our
data sets, both normalised to 1 at $t=0$ for better comparison. The flattening of
the slope for higher moments corresponds to a narrower spatial distribution of
partons within impact parameter space for $x \to 1$ \cite{Burkardt:2002hr}.

From the value of $A_{2,0}$ at $t=0$ we can determine the quark content and the
corresponding momentum fraction in the pion. For the data set shown in
Fig.~\ref{fig:moments} the renormalised value for the second moment is
$2\,A_{2,0}(t=0)=0.631(16)$, meaning that the quarks and antiquarks within the
pion carry about two thirds of the pion's momentum.

\section{Outlook}

The above analysis will be completed for all our lattices, and we have more data
for higher moments with up to three derivatives. The calculations include not
only the vector operators, but also the tensor combinations mentioned above. We
will furthermore be able to increase our statistics by using the symmetry
properties w.r.t.\ $\tsi$ of our three-point correlation function.

\section*{Acknowledgements}

The numerical calculations have been performed on the Hitachi SR8000 at LRZ
(Munich), on the Cray T3E at EPCC (Edinburgh) \cite{Allton:2001sk}, and on the
APEmille at NIC~/~DESY (Zeuthen). This work is supported in part by the DFG
(Forschergruppe Gitter-Hadronen-Ph\"anomenologie), by the EU Integrated
Infrastructure Initiative Hadron Physics under contract number
RII3-CT-2004-506078 and by the Helmholtz Association, contract number VH-NG-004.

\providecommand{\href}[2]{#2}\begingroup\raggedright\endgroup

\end{document}